\documentclass[prd,aps,twocolumn,showpacs,preprintnumbers,amsmath,amssymb]{revtex4}
\usepackage[dvips]{graphicx} 
\usepackage{amsmath} 
\usepackage{amssymb} 
 
\usepackage{graphicx}
\usepackage{dcolumn}
\usepackage{bm}
\def\be{\begin{equation}} 
\def\ee{\end{equation}} 
\def\bea{\begin{eqnarray}} 
\def\eea{\end{eqnarray}} 
 
\begin{document} 
 
 
\date{\today} 
 
\title{Cosmic Strings and Weak Gravitational Lensing} 
 
\author{Sergei Dyda$^{1,2)}$ \email[email: ]{}}

\author{Robert H. Brandenberger$^{1)}$ 
\email[email: ]{rhb@hep.physics.mcgill.ca}}
 
\affiliation{1) Department of Physics, McGill University, 
Montr\'eal, QC, H3A 2T8, Canada \\
2) Department of Physics, University of Toronto, Toronto, ON,
Canada}

\pacs{98.80.Cq} 
 
\begin{abstract} 

We study the deflection of light in the background of a
``wiggly" cosmic string, and investigate
whether it is possible to detect  cosmic strings by means
of weak gravitational lensing. For straight strings without
small-scale structure there are no signals. In the case of
strings with small-scale structure leading to a local
gravitational attractive force towards the string, there is
a small signal, namely a preferential elliptical distortion
of the shape of background galaxies in the direction corresponding
to the projection of the string onto the sky. The signal can be
statistically distinguished from the signal produced by a linear
distribution of black holes by employing an ellipticity axis distribution
statistic.
 
\end{abstract} 
 
\maketitle

\newcommand{\eq}[2]{\begin{equation}\label{#1}{#2}\end{equation}} 
 
\section{Introduction} 

There has been renewed interest in cosmic strings 
\cite{original} as a contributing
factor to cosmological structure formation. This resurgence of
interest is due firstly to the realization \cite{Rachel}
that in many supersymmetric particle physics models, cosmic strings are
formed after inflation, and thus contribute to but not completely
replace inflationary perturbations as the seeds for structure formation.
Secondly, it has recently been realized that models with cosmic superstrings 
\cite{Witten} may well be viable \cite{CMP}. They could, for example,
be generated as the remnant of brane annihilation processes in
brane inflation models \cite{Tye}, or they may play an important role
in inflationary models in warped backgrounds \cite{stringinflation}.
Cosmic superstrings may also be left behind after the initial Hagedorn
phase in string gas cosmology \cite{BV}, where they would add an additional
component to the spectrum of fluctuations produced by thermal string
gas fluctuations \cite{NBV}. 

Unlike in the original cosmic string models of structure formation
(see e.g. \cite{CSstructure} and \cite{CSreviews} for reviews), where it
was assumed that the strings were the sole source of structure formation,
in the context of the current models in which cosmic strings arise at
the end of inflation, the strings contribute only a small fraction of
the total power to the density fluctuations. The most stringent
constraints on the fraction of the power which cosmic strings can contribute 
come from measurements of the angular power spectrum of coordinates
cosmic microwave background (CMB)
anisotropies. A large fraction $f$ of the power being due to strings is 
inconsistent with the observed acoustic oscillations in the angular 
power spectrum. The best current limits on $f$ are \cite{CMBlimit}
$f < 10^{-1}$. Recent work \cite{ABB,Fraisse} points to the possibility
that statistical analyses searching for the line discontinuities \cite{KS}
in the microwave temperature maps produced by cosmic strings might lead to
even tighter limits.

Given their interest from the point of view of particle physics and
superstring theory, it is of great interest to develop statistics to
search for strings in observational data. There has been a substantial
amount of work on identifying distinctive signals for strings in
CMB temperature maps. In this paper we wish to take first steps at
exploring another avenue - weak gravitational lensing. Weak gravitational
lensing (see e.g. \cite{lensing} for a general review
of the applications of gravitational lensing to cosmology) is emerging 
as a powerful
tool in observational cosmology to search for the distribution of matter
in the universe. One of the advantages of weak gravitational lensing over,
for example, galaxy redshift surveys, is that light deflection depends on
the total mass, not just the luminous mass. As a consequence, cosmic
strings (which are also dark in the sense of not emitting light) also
lead to gravitational lensing. In fact, due to conical
structure of the metric of space-time in the plane perpendicular to a
long string \cite{deficit}, the specific lensing pattern produced
by a string can lead to interesting strong lensing patterns, e.g.
double images (see \cite{Esther,recent} for claims to have 
detected such events, claims which were subsequently shown to
be incorrect \cite{true}). In this short paper, we take a first step at
searching for weak lensing signals from strings. 

Cosmic strings \cite{original} are one-dimensional topological defects 
which arise during phase transitions in the very early universe. Since
they carry energy, they will lead to density fluctuations and
CMB anisotropies. 
Causality implies that the network of strings which forms during the
phase transition contains infinite strings.
Once formed in the early universe, the network of strings will approach a
``scaling solution'' which is characterized by of the order one infinite
string segment in each Hubble volume, and a distribution of cosmic string
loops which are the remnants of the previous evolution. In particular,
this implies that in any theory which admits cosmic strings, a network of
strings will be present at the current time. The network will consist
of a small number of strings crossing out entire Hubble volume (these are
commonly called the ``infinite'' strings). The mean curvature radius $R_c$
of these strings will be comparable to the Hubble radius. Thus, for
observations on smaller angular and distance scales, these strings
can be approximated as straight. In addition, there is a
distribution of string loops with radii $R$ between $R_c$ and a lower
scale set by the strength of the gravitational radiation which the loops
emit \cite{CSstructure}:
\be
G \mu R_c \, \le \, R \, \le \, R_c \, .
\ee
String loops with smaller radius live for less than one Hubble expansion
time. In the above,
$\mu$ is the mass per unit length of the string and $G$ is Newton's
gravitational constant

The distinctive signatures of strings in observational data are a
consequence of the specific form of the metric produced by a
cosmic string: space perpendicular to a cosmic string is a cone with
deficit angle given by \cite{deficit}
\be 
\alpha \, = \, 8 \pi G \mu \, .
\ee

In this paper we study the specific signature of this cosmic string metric
on weak gravitational lensing. We consider a cosmic string lens and
investigate the shape distortions this lens induces for a screen of
background galaxies as source objects.

To our knowledge, this is the first investigation of weak gravitational
lensing from cosmic strings. Strong lensing signatures (lines of
double images) have been studied extensity 
\cite{Vilenkin2,Hogan,Gott,Paczynski,Dyer,Vachaspati,Mack} theoretically,
and looked for in observational data sets \cite{Hindmarsh,Shirasaki}.

The outline of this paper is as follows: we first derive the deflection angle 
induced by a cosmic string with small scale structure. We then describe a 
gravitational lensing simulation based on this result, and propose a 
statistic which could be used to search for this type of effect in future 
weak lensing surveys.

\section{Cosmic String Lensing}

The deflection of light from distant sources by interposing masses is well 
understood in the context of General Relativity (GR). There are a variety
of lensing phenomena, beginning with strong lensing effects, in which case 
multiple images of a given source are observed. Other phenomena are
microlensing, an increase in the luminosity of a source as a lensing mass
passes close to the line of sight between observer and source, and weak 
lensing which is the shape distortion in the images of an extended source. 
Since all mass and not only visible matter lead to lensing, weak 
gravitational lensing \cite{Bartelmann} offers the hope of providing a tool
for mapping out the distribution of dark matter in the universe, amongst
other applications. Here, we will study weak lensing imprints of cosmic
strings, a specific class of dark objects.

In the usual theory of weak gravitational lensing \cite{Bartelmann}, 
lenses are characterized by their shear $\vec \gamma$ and their magnification 
$\mu$. These correspond, roughly speaking, to the 
distortion and magnification produced in the image. Since weak lensing 
effects are small and the source distribution is a priori unknown, searching 
for weak lensing relies on a statistical approach to discriminate 
between weakly lensed and un-lensed regions of the sky. The success of a 
weak lensing survey is therefore 
dependent on the statistic chosen to analyze the data.

In the following, we imagine a screen of background galaxies, modelled
as extended ellipsoidal objects. The light from these galaxies is lensed
by a foreground cosmic string.
   
Let us first consider the metric of a straight string without
small-scale structure (no ``wiggles") which is characterized
by a string tension $T$ which is equal in magnitude to the
mass per unit length $\mu$. In cylindrical coordinates, the metric of 
a such a straight string which we take to be oriented along the
z-axis is 
\begin{equation}
ds^2 \, = \, -dt^2 + dr^2 + dz^2 + (1-8G\mu )r^2 d\theta ^2,
\end{equation}
where G is the gravitational constant and $\mu$ is the mass per unit length
of the string \cite{deficit}. This is nothing but the
Minkowski metric with a deficit angle $\alpha = 8\pi G\mu$. Since
the metric is locally flat, no shape distortion is induced by 
such a string. 
 
Thus, we go on and consider a wiggly cosmic string (tension $T$ smaller
than the mass per unit length $\mu$) lying perpendicular to the 
optical axis, and derive the consequences for weak gravitational lensing in 
this space-time. We will use standard 
notation where $D_S$ is the distance from observer to source plane and 
$D_L$ the distance from observer to lens. 
For convenience, in cylindrical coordinates the z axis will lie along 
the direction of the string. The optical axis is 
chosen as the x direction. Consequently, the source and image planes will be 
parallel to the y-z plane, consistent with this choice of coordinates.

The deflection angle is defined as in the Schwarzschild case, namely as the 
deviation of a null geodesic from a source at 
infinity to an observer at infinity but propagating in the wiggly string 
metric. The deflection due to the small scale structure of the string 
is then defined as the deviation from null geodesics propagating in a 
straight string space-time.

\subsection{Null Geodesics}

For an infinite string lying along the z-axis the linearized wiggly string 
metric in cylindrical coordinates is
\begin{equation}
ds^{2} \, = \,
-(1+h_{00})dt^{2}+dr^{2}+(1-h_{00})dz^{2}+\omega^2 r^{2}d\theta^{2},
\end{equation} 
where 
\begin{eqnarray}
h_{00} \, &=& \, 4G(\mu-T)ln({r}/{r_{0}}) , \\
\omega^2 \, &\equiv& \, 1-4G(\mu+T) \, ,
\end{eqnarray}
G is the gravitational constant, and $r_{0}$ an integration 
constant \cite{deficit} which will not be relevant in our
analysis. 

Using the Lagrangian 
\begin{equation}
{\cal L} \, = \, \frac{1}{2}g_{\mu\nu}\dot{x}^{\mu}\dot{x}^{\nu}
\end{equation}
and the corresponding Euler-Lagrange equations we find the constants of
motion
\begin{eqnarray}
L \, &\equiv& \, \omega^2 r^{2}\dot{\theta}, \label{eq:constant3}  \\
K \, &\equiv& \, (1+h_{00})\dot{t}, \label{eq:constant1} \\ 
M \, &\equiv& (1-h_{00})\dot{z}. \label{eq:constant2}
\end{eqnarray}
Making use of (\ref{eq:constant3}), (\ref{eq:constant1}) and 
(\ref{eq:constant2}) we can 
find an effective Lagrangian
\begin{equation}
{\cal L}_{Eff} \, = \, \frac{1}{2}\left( - \frac{K^{2}}{(1+h_{00})} + \dot{r}^{2}  + \frac{M^{2}}{(1-h_{00})} + \frac{L^{2}}{\omega^2 r^{2}} \right).
\label{eq:effective}
\end{equation}

Using the condition $2L_{Eff}=0$ for null geodesics, we obtain the 
equation of motion in terms of the radial variable,
\begin{equation}
\dot{r}^{2} \, = \, \frac{K^{2}}{(1+h_{00})} - \frac{M^{2}}{(1-h_{00})} -  \frac{L^{2}}{\omega^2 r^{2}}.
\end{equation}
We now reparametrize in terms of $\theta$, defining 
$ r'\equiv {\dot{r}}/{\dot{\theta}}$. Using the change of variables 
$r\rightarrow {1}/{u}$ 
and differentiating once, (\ref{eq:effective}) can be rewritten 
\begin{equation}
u''= -\omega^2 u - \frac{2G(\mu -T)\omega^4}{L^2 u} \left[\frac{K^2}{(1 + h'_{00})^2} + \frac{M^2}{(1-h'_{00})^2} \right],
\label{eq:urt}
\end{equation}
where $h'_{00} = 4G(\mu-T)ln({u_{0}}/{u})$.

\subsection{Zeroth Order Solution (Straight String Contribution)}

Now proceed perturbatively by setting 
\begin{equation}
u \, \equiv \, u_{(0)} + u_{(1)} \, , 
\end{equation}
where the subscript denotes the power in our expansion parameter, namely
the magnitude of the small scale structure,  $G (\mu - T)$.
 
At zeroth order, (\ref{eq:urt}) reduces to 
\begin{equation}
u''_{(0)} \, = \, -\omega^2u_{(0)}.
\label{eq:zeroth}
\end{equation}
The initial conditions are specified by demanding $u(0) \equiv b^{-1}$, where 
b is the impact parameter. Solving (\ref{eq:zeroth}) yields 
\begin{equation}
u_{(0)}(\theta) \, = \, b^{-1}cos(\omega\theta).
\label{eq:zero}
\end{equation}

\subsection{First Order Solution (Contribution of Wiggles)}

Expanding (\ref{eq:urt}) to first order and substituting the zeroth order 
result (\ref{eq:zero}), we find 
\begin{equation}
u_{(1)}'' \, = \, -\omega^2u_{(1)} + \frac{C}{cos(\omega\theta)},
\label{eq:one}
\end{equation}
where 
\begin{equation}
C \, \equiv \, -\frac{2G(\mu - T)\omega^4 b (K^2 + M^2)}{L^{2}} \, . 
\end{equation}
This has the general solution 
\begin{eqnarray}
u_{(1)}(\theta) \, &=& \, 
C_1cos(\omega \theta) + C_2sin(\omega \theta) \\
&+& \frac{C}{\omega^2}(\omega \theta sin(\omega \theta) + cos(\omega \theta)ln(cos(\omega \theta))), \nonumber
\end{eqnarray}
where $C_1$,$C_2$ are integration constants. Applying the 
boundary conditions 
\begin{equation}
u_{(1)}(0) \, = \, u'_{(1)}(0) \, = \, 0  
\end{equation}
at the point of nearest approach we find
\begin{equation}
u_{(1)}(\theta) \, = \, \frac{C}{\omega^2}(\omega \theta sin(\omega \theta) + cos(\omega \theta)ln(cos(\omega \theta))).
\end{equation}

\subsection{Deflection Angle}

To calculate the deflection angle we consider the deviation 
\begin{equation}
\Delta \theta \, \equiv \, \frac{\pi}{2} - \theta
\end{equation} 
at small values of u ($r \rightarrow \infty$). This quantity is 
then doubled since the deflection angle is defined as the deviation 
due to light propagating from $-\infty \rightarrow \infty$.

First, consider the deflection angle for a straight string without
small-scale structure, i.e. working with the unperturbed solution
$u_{(0)}$. Setting $u_{(0)}=0$ and expanding by using the fact that 
$\Delta \theta << 1$ we find a deflection angle 
\begin{equation}
\alpha_{s} \, = \, 2\Delta \theta = 2\pi G(\mu +T) \, ,
\label{eq:sangle}
\end{equation}
the standard result.

Next, we move on to the case of the wiggly string.
As above, we set $u = u_{(0)} + u_{(1)} = 0$ for 
$\theta \approx \frac{\pi}{2}$ and 
solve for $\Delta \theta $ (defined above). In this approximation 
$sin(\omega \theta) \approx 1$ and 
$cos(\omega \theta) \approx \omega \Delta \theta$. Hence 
$u_{(0)} = b^{-1}\omega \Delta \theta$ 
and the leading term for the first order perturbation is 
\begin{equation}
u_{(1)} \, \approx \, \frac{C \pi}{2 \omega} \, . 
\end{equation}
This yields a deviation
\begin{equation}
\Delta \theta \, = \, \frac{-Cb\pi}{2\omega ^2},
\end{equation} 
and a deflection angle
\begin{equation}
\alpha_w \, = \, 2 \Delta \theta \approx -2G(\mu -T)\omega^2 b^2 \pi 
\left( \frac{K^2 + M^2}{L^2} \right) .
\end{equation}
Using the null condition and evaluating the constants of the motion at 
$\theta = 0$ yields the deflection angle due to string structure
\begin{equation}
\alpha_w = \frac{2\pi G(\mu -T)}{\omega^2}\left( 2\left(\frac{\dot{z}}{b\dot{\theta}}\right)^2 + 1 \right).
\label{ialpha}
\end{equation}
This is a constant deflection term, plus a term that is position dependent 
which may lead to some shearing.

\subsection{Initial Conditions}

\begin{figure}
\centering
\includegraphics[scale=0.5]{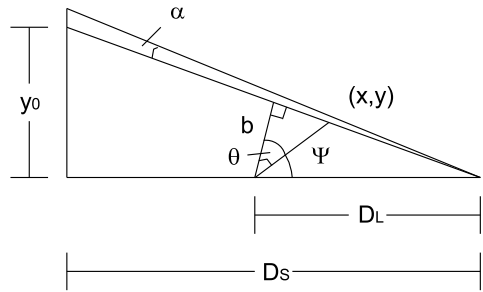}
\caption{The geometry of the problem, where the azimuthal angle 
$\theta = 0$ at the point of closest approach.}
\label{fig:string}
\end{figure}

We evaluate the initial conditions for the zeroth order geodesic, and use 
these to calculate the first order deflection. At zeroth order, geodesics 
are straight lines so we will work in Cartesian coordinates. Parametrizing 
the geodesic for $0 < \lambda < 1$, where $\lambda = 0$ corresponds to the 
source position and $\lambda = 1$ corresponds to the observer yields 
\begin{align}
x = (D_L - D_S) + \lambda D_S, \\ 
y = y_0 - \lambda y_0, \\
z = z_0 - \lambda z_0.
\label{cart}
\end{align} 
{F}rom the geometry (Fig. \ref{fig:string}) we find
\begin{equation}
\theta \, = \, \Psi - tan^{-1}\left( \frac{y}{x} \right),
\end{equation}
where $\Psi(b,D_L)$. Differentiation with respect to $\lambda$ yields
\begin{equation}
\dot{\theta} \, = \, \frac{y\dot{x} - \dot{y}x}{x^2 + y^2}.
\end{equation}
Substituting (\ref{cart}) and evaluating at $\theta =0$ yields 
\begin{equation}
\dot{\theta} \, = \, \frac{D_S^2 + y_0^2}{D_Ly_0}.
\end{equation}
Substituting into (\ref{ialpha}) yields 
\begin{equation}
\alpha_w \, = \, \frac{2\pi G(\mu -T)}{\omega^2}\left( \frac{2 z_0^2}{(D_S^2 + y_0^2)} + 1 \right).
\label{alpha}
\end{equation}

\section{Results}

\subsection{Ellipticity Axis Distribution as a Weak Lensing Statistic}

In order to search for cosmic string lensing we propose using the 
ellipticity axis distribution (EAD) defined as follows. Consider an
elliptical object in the image plane with ellipticity 
$\epsilon \equiv b/a$ where b, a are the semi-minor and semi-major 
axes respectively. Define the ellipticity vector $\vec \epsilon$ as a 
unit vector lying in direction of the  major axis of the ellipse. 
Now define the ellipticity axis angle (EAA) $\phi$ via
\begin{equation}
\phi \, = \, cos^{-1}(\vec \epsilon \cdot \vec z),
\label{eq:ellipticity}
\end{equation} 
where we have chosen to measure the angle with respect to the unit 
vector $\vec z$. The ellipticity axis angle is hence a measure of 
the orientation of an ellipse's major axis 
in the image plane. We will call the distribution of EAA for a set of 
objects the ellipticity axis distribution (EAD). 

In an un-lensed region, one would expect a uniform EAD due to isotropy. 
However, we will show that cosmic string lensing is predicted to 
produce a peak in the EAD for a subset of sources. 

\subsection{Shearing of a nearly circular source}

We now examine how (\ref{alpha}) leads to a specific signature in the EAD 
of nearly circular sources. We will call a source nearly circular if it is
to first order circular in shape (i.e. $z^2 + y^2 \approx r^2$), but unlike
a circle has a well defined EAA. The radius of this source is $r$.

Consider such a source in Cartesian coordinates centered at $(D_S,0,D_S)$.
Points on the surface of the object have coordinates 
\begin{equation}
(D_S,y,D_S + z) \, = \, (D_S,r sin(\phi),D_S + r cos(\phi)),  
\end{equation} 
for $0< \phi < 2\pi$. 
We want to find the point where the radius change of
the image is maximized. 
Expanding (\ref{alpha}) to first order in r we find the non-constant part
of the deflection to be
\begin{equation}
\alpha_{\Delta} \, = \, k r cos(\phi),
\end{equation}  
where $k = \frac{8 \pi G (\mu - T)}{\omega^2 D_S}$. The radius of the 
image point is then approximately $R^2 = z^2 + (y + \alpha_\Delta D_S)^2$
where we have used the fact that the z coordinate is unaffected by the 
lens and the non-constant shift in the y direction is 
$\Delta y \approx \alpha_\Delta D_S$. Expanding to first order in 
$\alpha_\Delta$ yields
\begin{equation}
R^2 \, = \, r^2 + 2kr^2D_S sin(\phi)cos(\phi),
\end{equation}
where we have used the nearly circular property of the source. 
Differentiating we find 
\begin{equation}
R\frac{dR}{d\phi} = k(cos^2(\phi) - sin^2(\phi))r^2.
\end{equation}
Hence R is extremized at $\phi= \pi/4$, and it is easy to check that this 
is a maximum. For nearly circular sources this admits the possibility that
the EAD will have a sharp peak because the EAA of the images will 
preferentially be at $\phi = \pi/4$ since this is the angle at which the
greatest shearing occurs. In contrast, an object centered at the position 
$(D_S,0,-D_S)$ would experience a shear which would lead to an EAD peaked at 
$\phi = -\pi/4$. 

\subsection{Lensing simulation}

The goal was to determine the threshold ellipticity of galaxies which 
would produce a detectable signal in the EAD. We begin with a short
description of the code.

The source plane is created by forming a regularly spaced lattice and 
placing a galaxy centered at each lattice 
point. The lensing map is applied to each galaxy in the source plane 
to produce the image plane. 
The ellipticity angle of each galaxy is measured by identifying the 
major axis and using (\ref{eq:ellipticity}). These results are compiled 
to produce the EAD. 

We fixed the small parameter responsible for the effect to be 
$4 \pi G (\mu-T) = 10^{-6}$, a value which corresponds to the
upper range allowed by the current constraints on $\mu$.
The source plane was fixed at a distance of redshift $z = 0.1$, 
the typical redshift of galaxies in current large-scale
galaxy redshift surveys, and the 
lens plane at a redshift of $z = 0.05$. The mean galaxy size was 
chosen to be $r = 30 000$pc, similar to the size of our Milky Way
galaxy.

\begin{figure}
\centering
\includegraphics[scale=0.65]{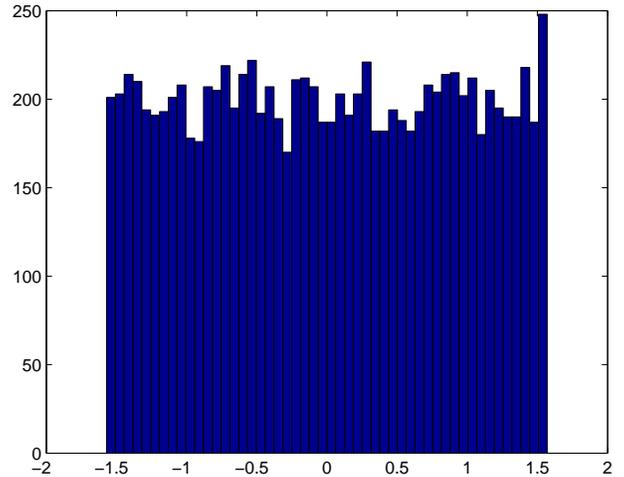}
\caption{EAD of the source plane corresponding to a random choice of
ellipticity axes in the interval $[-\pi/2,\pi/2]$ (uniform measure).}
\label{fig:EAD2}
\end{figure}

Galaxies have an intrinsic ellipticity which will act as
background noise to our analysis. We have chosen the ellipticity 
magnitude $\epsilon$ to be randomly chosen 
in the range $\epsilon \in [t,1]$ according to a uniformly
distributed measure, where $t \approx 1$, i.e.
nearly spherical objects, needed to be taken in order to
be able to obtain a weak lensing signature from strings. The
axis of each source object was 
randomly chosen in the range $\phi \in [-\pi/2,\pi/2]$, again
according to a uniform measure 
(Fig. \ref{fig:EAD2}). The cutoff ellipticity parameter $t$ was 
increased until a detectable signal was achieved in the EAD of the
objects in the image plane. The number of objects in the source
plane was taken to be 10,000. 
 
\begin{figure}
\centering
\includegraphics[scale=0.65]{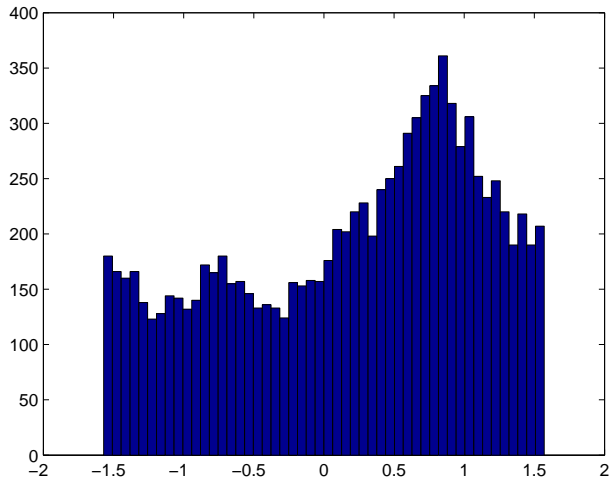}
\caption{EAD of the image plane for sources far from the optical axis 
($z \approx D_S$) and EAA randomly chosen in the range [0.995,1]
(uniform measure). As predicted, a sharp peak appears at $\phi \approx \pi/4$.}
\label{fig:EAD}
\end{figure}

By varying the parameter t, it 
was found that a detectable signal in the EAD could be measured for 
$t = 0.995$, for objects at a distance $z \approx D_S$ 
(Fig. \ref{fig:EAD}). Decreasing the strength of the small scale 
structure led to a corresponding increase in t, the threshold 
ellipticity. At fixed t, decreasing the value of z caused the signal 
to gradually vanish.

\subsection{Search Strategy} 

To search for this effect one might first divide the sky into a grid 
and find the EAD of sources with ellipticity $\epsilon \in [t,1]$ in 
each grid square. Then determine the peak in the EAD of each grid 
square and search for a linear distribution of grid squares where the 
EAD peak varies from $-\pi/4$ to $\pi/4$ across the distribution. 
This would signal the presence of a cosmic string stretching across 
the linear distribution of grid squares. 

Searching for this lensing effect can therefore be used to set bounds on 
the small scale structure of any potential cosmic strings in our Hubble 
volume. In the optimistic case that this effect is observed, it would 
provide a motivation and starting point to search for the unique strong
lensing signature of a straight cosmic string. 

It is important to point out that a combination of ordinary Schwarzschild 
lenses could in principle produce a similar EAD signature. Since
the lensing mass is not uniformly distributed along the z direction,
the lensing angle distribution would be less regular, thus allowing
a discrimination between the signals from a string and from a line
of Schwarzschild lenses. However, because 
of the huge degeneracy of this problem, it seems premature to attempt an 
analysis without actual data. We conclude that this lensing signature be 
used as a first step in subsequent analyses of weak lensing surveys, and 
that more rigorous analysis, in particular searching for the strong lensing 
effect caused by the deficit cone, be used to pinpoint the exact location 
of the string.      

\section{Conclusions}
 
In this work we have studied the gravitational lensing by a straight
cosmic string containing small-scale structure which leads to a string
tension which is less than the mass per unit length of the string,
and thus induces a net gravitational force towards the
string which test particles feel.

Next, we studied potential weak lensing signatures of such wiggly
strings. We found a shape distortion which is proportional to
$G (\mu - T)$ but depends on the location of the source relative
to the line of sight between the observer and the string. Only
objects which are displaced in direction of the string relative
to the line of sight give rise to a shape distortion. The shape
distortion increases linearly as the distance in string direction
increases.

The specific distribution of the weak lensing distortion in
the image plane in principle can be used to provide a signature
for cosmic strings. However, in practise the signal appears
to be too small to be useful.
Even for optimal choices of parameters (galaxy redshifts comparable
to those of galaxies in the largest current redshift surveys),
distance from the line of sight comparable to the distance of
the string from the observer, large amount of small scale 
structure and string tension close to the current observational
bounds, the intrinsic ellipticity of the background galaxies
needs to be very close to one (i.e. the shape needs to be
very close to spherical) in order for the cosmic string
signal to stand out.

\begin{acknowledgments} 
 
This work is supported in part by a NSERC Discovery Grant, by funds from
the Canada Research Chair Program, and by a FQRNT Team Grant.

\end{acknowledgments}


\begin{thebibliography}{99} 
 
\bibitem{original}
T.~W.~B.~Kibble,
  ``Topology Of Cosmic Domains And Strings,''
  J.\ Phys.\ A  {\bf 9}, 1387 (1976);\\
Y.~B.~Zeldovich,
  ``Cosmological fluctuations produced near a singularity,''
  Mon.\ Not.\ Roy.\ Astron.\ Soc.\  {\bf 192}, 663 (1980);\\
A.~Vilenkin,
  ``Cosmological Density Fluctuations Produced By Vacuum Strings,''
  Phys.\ Rev.\ Lett.\  {\bf 46}, 1169 (1981)
  [Erratum-ibid.\  {\bf 46}, 1496 (1981)].

\bibitem{Rachel}
R.~Jeannerot,
  ``A Supersymmetric SO(10) Model with Inflation and Cosmic Strings,''
  Phys.\ Rev.\  D {\bf 53}, 5426 (1996)
  [arXiv:hep-ph/9509365];\\
R.~Jeannerot, J.~Rocher and M.~Sakellariadou,
  ``How generic is cosmic string formation in SUSY GUTs,''
  Phys.\ Rev.\  D {\bf 68}, 103514 (2003)
  [arXiv:hep-ph/0308134].

\bibitem{Witten}
E.~Witten,
  ``Cosmic Superstrings,''
  Phys.\ Lett.\  B {\bf 153}, 243 (1985).

\bibitem{CMP}
E.~J.~Copeland, R.~C.~Myers and J.~Polchinski,
  ``Cosmic F- and D-strings,''
  JHEP {\bf 0406}, 013 (2004)
  [arXiv:hep-th/0312067].

\bibitem{Tye}
S.~Sarangi and S.~H.~H.~Tye,
  ``Cosmic string production towards the end of brane inflation,''
  Phys.\ Lett.\  B {\bf 536}, 185 (2002)
  [arXiv:hep-th/0204074].

\bibitem{stringinflation}
C.~P.~Burgess,
  ``Inflatable string theory?,''
  Pramana {\bf 63}, 1269 (2004)
  [arXiv:hep-th/0408037];\\
J.~M.~Cline,
  ``Inflation from string theory,''
  arXiv:hep-th/0501179;\\
A.~Linde,
  ``Inflation and string cosmology,''
  eConf {\bf C040802}, L024 (2004)
  [arXiv:hep-th/0503195].

\bibitem{BV}
R.~H.~Brandenberger and C.~Vafa, 
  ``Superstrings In The Early Universe,'' 
  Nucl.\ Phys.\ B {\bf 316}, 391 (1989). 

\bibitem{NBV}
A.~Nayeri, R.~H.~Brandenberger and C.~Vafa, 
  ``Producing a scale-invariant spectrum of perturbations in a Hagedorn phase 
  of string cosmology,''
 Phys.\ Rev.\ Lett.\  {\bf 97}, 021302 (2006)   [arXiv:hep-th/0511140];\\
 R.~H.~Brandenberger, A.~Nayeri, S.~P.~Patil and C.~Vafa,
  ``String gas cosmology and structure formation,''
  arXiv:hep-th/0608121.

\bibitem{CSstructure}
N.~Turok and R.~H.~Brandenberger,
  ``Cosmic Strings And The Formation Of Galaxies And Clusters Of Galaxies,''
  Phys.\ Rev.\ D {\bf 33}, 2175 (1986);\\
H. Sato, ``Galaxy Formation by Cosmic Strings,''
  Prog. Theor. Phys.\  {\bf 75}, 1342 (1986);\\
A. Stebbins, ``Cosmic Strings and Cold Matter'',
  Ap. J. (Lett.) {\bf 303}, L21 (1986).

\bibitem{CSreviews}
A. Vilenkin and E.P.S. Shellard;
\textit{Cosmic Strings and Other Topological Defects},
(Cambridge Univ. Press, Cambridge, 1994);\\
M.~B.~Hindmarsh and T.~W.~Kibble,
``Cosmic strings,''
Rept.\ Prog.\ Phys.\  {\bf 58}, 477 (1995)
[arXiv:hep-ph/9411342];\\
R.~H.~Brandenberger,
``Topological defects and structure formation,''
Int.\ J.\ Mod.\ Phys.\ A {\bf 9}, 2117 (1994)
[arXiv:astro-ph/9310041].

\bibitem{CMBlimit}
L.~Pogosian, S.~H.~H.~Tye, I.~Wasserman and M.~Wyman,
  ``Observational constraints on cosmic string production during brane
  inflation,''
  Phys.\ Rev.\  D {\bf 68}, 023506 (2003)
  [Erratum-ibid.\  D {\bf 73}, 089904 (2006)]
  [arXiv:hep-th/0304188];\\
M.~Wyman, L.~Pogosian and I.~Wasserman,
  ``Bounds on cosmic strings from WMAP and SDSS,''
  Phys.\ Rev.\  D {\bf 72}, 023513 (2005)
  [Erratum-ibid.\  D {\bf 73}, 089905 (2006)]
  [arXiv:astro-ph/0503364];\\
A.~A.~Fraisse,  
  ``Constraints on topological defects energy density from first year WMAP     
  results,''  
  arXiv:astro-ph/0503402.  

\bibitem{ABB}
S.~Amsel, J.~Berger and R.~H.~Brandenberger,  
  ``Detecting Cosmic Strings in the CMB with the Canny Algorithm,''  
  arXiv:0709.0982 [astro-ph].  

\bibitem{Fraisse}
A.~A.~Fraisse, C.~Ringeval, D.~N.~Spergel and F.~R.~Bouchet,  
  ``Small-Angle CMB Temperature Anisotropies Induced by Cosmic Strings,''  
  arXiv:0708.1162 [astro-ph].  

\bibitem{KS}
N.~Kaiser and A.~Stebbins,
  ``Microwave Anisotropy Due To Cosmic Strings,''
  Nature {\bf 310}, 391 (1984).

\bibitem{lensing}
P. Schneider, J. Ehlers and E. Falco, Gravitational Lenses 
  (Springer, New York, 1992);\\
P.~Schneider,  ``Cosmological Applications of Gravitational Lensing,''   
  arXiv:astro-ph/9512047.  

\bibitem{deficit}
A.~Vilenkin,
  ``Gravitational Field Of Vacuum Domain Walls And Strings,''
  Phys.\ Rev.\  D {\bf 23}, 852 (1981).

\bibitem{Esther}
L. Cowie and E. Hu,
  ``The formation of families of twin galaxies by string loops, ''
  Astrophys. J. (Lett.) {\bf 318}, L33 (1987).

\bibitem{recent}
M.~Sazhin {\it et al.},  
  ``CSL-1: a chance projection effect or serendipitous discovery of a   
  gravitational lens induced by a cosmic string?,''  
  Mon.\ Not.\ Roy.\ Astron.\ Soc.\  {\bf 343}, 353 (2003)  
  [arXiv:astro-ph/0302547].  

\bibitem{true}
M.~V.~Sazhin, M.~Capaccioli, G.~Longo, M.~Paolillo and O.~S.~Khovanskaya,   
  ``The true nature of CSL-1,''  
  arXiv:astro-ph/0601494.  

\bibitem{Vilenkin2}
A.~Vilenkin,
  ``Cosmic Strings As Gravitational Lenses,''
  Astrophys.\ J.\  {\bf 282}, L51 (1984).

\bibitem{Hogan}
C. Hogan and R. Narayan, 
  ``Gravitational lensing by cosmic strings",
  MNRAS {\bf 211}, 575 (1984).

\bibitem{Gott}
J.~R.~I.~Gott,
  ``Gravitational lensing effects of vacuum strings: Exact solutions,''
  Astrophys.\ J.\  {\bf 288}, 422 (1985).

\bibitem{Paczynski}
B. Paczynski, 
  ``Will cosmic strings be discovered using the Space Telescope?",
  Nature {\bf 319}, 567 (1986).

\bibitem{Dyer}
C.~C.~Dyer and F.~R.~Marleau,
  ``Complete model of a selfgravitating cosmic string. 1: A New class of exact
  solutions and gravitational lensing,''
  Phys.\ Rev.\  D {\bf 52}, 5588 (1995)
  [arXiv:astro-ph/9411087].

\bibitem{Vachaspati}
A.~A.~de Laix and T.~Vachaspati,
  ``Gravitational lensing by cosmic string loops,''
  Phys.\ Rev.\  D {\bf 54}, 4780 (1996)
  [arXiv:astro-ph/9605171].

\bibitem{Mack}
K.~J.~Mack, D.~H.~Wesley and L.~J.~King,
  ``Observing cosmic string loops with gravitational lensing surveys,''
  arXiv:astro-ph/0702648.

\bibitem{Hindmarsh}
M.~Hindmarsh,
  ``Searching For Cosmic Strings,''
{\it  In *Cambridge 1989, Proceedings, The formation and evolution of cosmic strings* 527-542. (see HIGH ENERGY PHYSICS INDEX 29 (1991) No.
7720)}

\bibitem{Shirasaki}
Y.~Shirasaki {\it et al.},
  ``Searching for a long cosmic string through the gravitational lensing
  effect,''
  arXiv:astro-ph/0305353.

\bibitem{Bartelmann}
M.~Bartelmann and P.~Schneider,  
  ``Weak Gravitational Lensing,''  
  Phys.\ Rept.\  {\bf 340}, 291 (2001)  
  [arXiv:astro-ph/9912508].  

\end{thebibliography}
\end{document}